\documentclass[prd,reprint,floatfix]{revtex4-1}   

\usepackage{amsmath}    
\usepackage{graphicx}   

\begin{document}

\title{Differences in Quasi-Elastic Cross-Sections of Muon and Electron Neutrinos}
\author{Melanie Day}
 \affiliation{University of Rochester, Department of Physics and Astronomy, Rochester, NY 14627 USA}
\author{Kevin S.\ McFarland}
 \affiliation{University of Rochester, Department of Physics and Astronomy, Rochester, NY 14627 USA\\and Fermi National Accelerator Laboratory, Batavia, IL 60510 USA}
\date{\today}

\begin{abstract}
Accelerator neutrino oscillation experiments seek to make precision
measurements of the neutrino flavor oscillations
$\stackrel{(-)}{\nu}_\mu\to\stackrel{(-)}{\nu}_e$ in order to
determine the mass hierarchy of neutrinos and to search for CP
violation in neutrino oscillations.  These experiments are currently
performed with beams of muon neutrinos at energies near 1~GeV where
the charged-current quasi-elastic interactions $\nu_\ell n\to\ell^-p$
and $\bar{\nu}_\ell p\to\ell^+n$ dominate the signal reactions.  We
examine the difference between the quasi-elastic
cross-sections for muon and electron neutrinos and anti-neutrinos and
estimate the uncertainties on these differences.

\end{abstract}

\maketitle

\section{Introduction}

Since the invention of neutrino beams at accelerators and the
consequent discovery of the two flavors of
neutrinos\cite{Lederman-Schwartz-Steinberger}, the reactions
$\nu_\ell n\to \ell^- p$ and $\bar{\nu}_\ell p \to \ell^+ n$, which are
the dominant reactions of muon and electron neutrinos with energies from 
$200$~MeV to $2$~GeV, have played
an important role in studies of neutrino flavor.  These
charged-current quasi-elastic (CCQE) interactions are important not
only because they identify the flavor of the neutrino through the
production of the lepton in the final state, but also because the two
body kinematics permit a measurement of the neutrino energy from only
the observation of the final state lepton.

Accelerator neutrino experiments like T2K\cite{T2K1,T2K2}, 
NOvA\cite{NOvA} and a number of proposed
experiments seek to make precision measurements of the neutrino flavor
oscillations $\stackrel{(-)}{\nu}_\mu\to\stackrel{(-)}{\nu}_e$ or
$\stackrel{(-)}{\nu}_e\to\stackrel{(-)}{\nu}_\mu$ in order to
determine the mass hierarchy of neutrinos and to search for CP
violation in neutrino oscillations.  Uncertainties on differences
between these cross-sections for muon and electron neutrinos will
contribute to experimental uncertainties in these flavor oscillation
measurements.

Interactions of the charged-current with fundamental
fermions, such as $\nu_\ell d \to \ell^- u$, have no
uncertainties in the differences between the reactions for muon and
electron neutrino interactions because the weak interaction is
experimentally observed to be flavor universal.  In particular, the
effect of the final state lepton mass in this two body reaction of
fundamental fermions can be unambiguously calculated.  

One such calculable difference occurs because of radiative corrections
to the tree-level CCQE process.  Radiative corrections from a particle
of mass $m$ in a process with momentum transfer $Q$ are of order
$\frac{\alpha}{\pi}\log\frac{Q}{m}$, which implies a significant
difference due to the mass of the final state
lepton\cite{De_Rujula:1979jj}.  Although this effect is
calculable, it is not accounted for in neutrino
interaction generators used in recent analysis of experimental data,
such as GENIE\cite{GENIE}, NEUT\cite{NEUT,NEUT2} and
NUANCE\cite{NUANCE}.

There are, however, cross-section differences due to lepton mass 
which cannot be calculated
from first principles with current theoretical tools.  The
presence of the target quarks inside a strongly bound nucleon lead to
a series of {\em a priori} unknown form factors in the nucleon level
description of the reaction, e.g., $\nu_\ell n\to\ell^-p$.  It is the
uncertainties on these form factors combined with the alteration of
the kinematics due to lepton mass that leads to uncertainties, and
that is the focus of the results of this paper.

There is also a modification of the reaction cross-sections due to
the effects of the nucleus in which the target nucleons are bound.
The model incorporated in GENIE\cite{GENIE}, NEUT\cite{NEUT,NEUT2} and
NUANCE\cite{NUANCE} is a relativistic Fermi gas
model\cite{Smith-Moniz,Bodek-Ritchie} which provides a distribution
of nucleon kinematics inside the nucleus.  A more sophisticated
description from a nuclear spectral function model\cite{Benhar}
is implemented in the NuWro generator\cite{NuWro}.  We do
not consider the effect of the nucleus in this work, although it may
be important in the relative weighting of nucleon kinematics at low
energy.  However, this work provides a blueprint for studying the
effect of the final state lepton mass in different nuclear models.

\section{Nucleon Form Factors}

The cross section for quasi-elastic scattering of neutrinos at
energies relevant for oscillation experiments may be calculated from
the Fermi theory of weak interactions with the effective Lagrangian,
\begin{equation}
\label{eq:effLagranian}
{\cal L}_{\rm\textstyle eff}=\frac{G_F}{\sqrt{2}}\left( J_{(\ell)\lambda}^\dagger J_{(H)}^\lambda + {\rm\textstyle Hermitian~conjugate}\right) ,
\end{equation}
where $G_F$ is the Fermi constant and the $J$ are the leptonic and
hadronic currents.  The form of the leptonic current is specified by
the theory to be
\begin{equation}
\label{eq:leptonicCurrent}
J_{(\ell)\lambda} = \bar{\psi}_{\ell}\gamma_\lambda(1-\gamma_5)\psi_{\nu_\ell},
\end{equation}
because the leptons are fundamental fermions.  However, as mentioned above
the hadronic current for quasi-elastic scattering
depends on unknown form factors of
the nucleons.  The hadronic current can be decomposed into 
vector and axial components,
\begin{equation}
\label{eq:hadronicCurrent}
J_{(H)}^{\lambda} = J_V^{\lambda} + J_A^{\lambda}.
\end{equation}
$J_V$ contains three terms related to the vector form factors $F_V^1$,
$F_V^2$ and $F_V^3$, and $J_A$ contains three terms related to
the axial form factors $F_A$, $F_A^3$ and $F_p$.  A description the the
bilinear covariant structure of the currents is given in several standard texts and review papers\cite{marshak,lsmith,wilkinson}.  We follow most closely the
notation of Ref.~\citenum{lsmith}. 

From the effective Lagrangian of Eq.~\ref{eq:effLagranian} and currents above in Eqs.~\ref{eq:leptonicCurrent} and \ref{eq:hadronicCurrent}, the quasi-elastic cross
section on free nucleons is:
\begin{eqnarray}
\label{eq:lsshort}
\frac{d\sigma}{dQ^2}( ^{\nu n \rightarrow l^- p} _{\overline{\nu} p \rightarrow l^+ n}) &=& 
\left[A(Q^2) \mp B(Q^2)\frac{s-u}{M^2}
+C(Q^2)\frac{(s-u)^2}{M^4}\right] \nonumber\\
&&\times\frac{M^2 G_F^2 \cos^2 \theta_{c}}{8 \pi E_{\nu}^2}
\end{eqnarray}
where the invariant $Q^2=-q^2$ and $q$ is the four momentum
transfer from the leptonic to hadronic system, $M$ is the mass of the
nucleon, $\theta_c$ is the Cabibbo
angle, and $E_{\nu}$ is the neutrino energy in the lab.  The
combination of Mandelstam invariants $s$ and $u$ can be written as,
\begin{equation}
\label{eq:su}
s-u = 4 M E_{\nu} -Q^2 -m^2,
\end{equation}
where $m$ is the mass of the final state lepton. The functions
A($Q^2$), B($Q^2$) and C($Q^2$) depend on the nucleon form factors 
and $\xi$, the difference between the anomalous
magnetic moment of the proton and the neutron:
\begin{widetext}
\begin{eqnarray}
\label{eq:Afunc}
A(Q^2) &=& \frac{m^2+Q^2}{4 M^2}\left[ \left( 4+ \frac{Q^2}{M^2} \right) \vert F_{A} \vert^2 - \left( 4- \frac{Q^2}{M^2} \right) \vert F_{V}^1 \vert ^2 \right. 
 + \frac{Q^2}{M^2} \xi \vert F_{V}^2 \vert ^2 \left( 1 - \frac{Q^2}{4 M^2} \right)  + \frac{4 Q^2 Re F_V^{1*} \xi F_V^{2}}{M^2} \nonumber \\
&&  - \frac{Q^2}{M^2} \left( 4+ \frac{Q^2}{M^2} \right) \vert F_A^3
  \vert ^2   - \frac{m^2}{M^2} \left( \left.  \vert F_V^1 + \xi F_V^2  \vert
  ^2 
   + \vert F_A + 2 F_P \vert ^2  - \left(
   4+ \frac{Q^2}{M^2} \right) \left( \vert F_V^3 \vert ^2 + \vert F_P
   \vert ^2 \right) \right) \right] ,\\
\label{eq:Bfunc}
B(Q^2) 
&=& \frac{Q^2}{M^2} Re F_A^* \left( F_V^1 + \xi F_V^2\right)
- \frac{m^2}{M^2} Re \left[ \left( F_V^1 -\frac{Q^2}{4 M^2} \xi F_V^2 \right) ^* F_V^3 \right. 
\left. - \left( F_A - \frac{Q^2 F_P}{2 M^2} \right)^* F_A^3 \right]
{\rm\textstyle and}\\
\label{eq:Cfunc}
C(Q^2) &=& \frac{1}{4} \left( \vert F_A \vert ^2 + \vert F_V^1 \vert ^2
+ \frac{Q^2}{M^2} \left| \frac{\xi F_V^2}{2} \right| ^2 +
\frac{Q^2}{M^2} \vert F_A^3 \vert ^2 \right) .
\end{eqnarray}
\end{widetext}
Note that the form factors themselves are functions of $Q^2$ in
Eqs.~\ref{eq:Afunc}--\ref{eq:Cfunc}.
 
$F_V^1$ and $F_V^2$ are the vector and $F_A$ and $F_P$ the axial form
factors of the first class currents. First class currents conserve 
both time and charge symmetry.  In addition, first class
vector currents commute with the G-parity operator while first class
axial currents anti-commute with it\cite{marshak}.  The terms
associated with $F_V^1$ and $F_A$ are considered the leading terms in
the hadron current since they have no dependence on the four-momentum
transfer, excepting that of the form factors, and they are not suppressed
by powers of the final state lepton mass as $F_P$ is.

Vector elastic form factors are precisely known at $Q^2=0$ from the
nucleon electric charges and magnetic moments and are precisely
measured over a wide range of $Q^2$ in charged-lepton elastic
scattering from protons and deuterium.  The axial nucleon form factor
at zero $Q^2$ is precisely measured in studies of neutron beta
decay.  At higher $Q^2$, much of our knowledge of the axial form
factors comes from muon neutrino quasi-elastic scattering
measurements.  For $Q^2\stackrel{<}{\sim}1$~(GeV/c)$^2$, the vector
form factors and the axial form factors are observed to follow a
dipole form, that is
\begin{equation}
\label{eq:dipole}
F(Q^2)\propto F(0)/(1+Q^2/C^2)^2
\end{equation}
where the constant $C$ is often expressed as a mass of the same order
of magnitude as the mass of the nucleon.  At high $Q^2$ the vector
form factors do not follow the dipole structure\cite{BBBA}.  The
neutrino scattering data 
contains conflicting results among different
experiments\cite{Bodek:2007vi,Lyubushkin:2008pe,AlcarazAunion:2009ku,Dorman:2009zz,AguilarArevalo:2010zc}
which limit our ability to effectively use that information to constrain the axial form factor.  Pion
electroproduction experiments\cite{Choi:1993vt,Liesenfeld:1999mv} have
also measured the axial form factor at $Q^2<$~0.2~(GeV/c)$^2$.

The form factor $F_P$
is determined from PCAC which, under minimal assumptions,
states that\cite{adler}:
\begin{equation}
\label{eq:PCAC}
\delta_{\mu} J_A = C \phi
\end{equation}
where $\phi$ is the renormalized field operator that creates the
$\pi^+$ and $C$ is a constant which may be computed at $Q^2=0$. 
PCAC gives the following relation between
$F_P$ and the pion nucleon form factor, $g_{\pi NN}$,
\begin{eqnarray}
\label{eqn:PCACderiv}
F_P(Q^2)&=&\frac{2 M^2 F_A(0)}{Q^2} \nonumber\\
&&\times \left( \frac{F_A(Q^2)}{F_A(0)} -
\frac{g_{\pi NN}(Q^2)}{g_{\pi NN}(0)}\frac{1}{(1 +
  \frac{Q^2}{M_{\pi}^2})}\right) ,
\end{eqnarray}
where $M_{\pi}$ is the pion mass. 
The Goldberger-Treiman relation\cite{goldtrei} predicts
\begin{equation}
\label{eq:gt}
g_{\pi NN}(Q^2)F_{\pi} = F_A(Q^2)M,
\end{equation}
where $F_{\pi}$ is
the pion decay constant. 
Under the assumption that the Goldberger-Treiman relation holds for
all values of $Q^2$, then $F_P$ is
\begin{equation}
\label{eqn:fp}
F_P(Q^2) = \frac{2 M^2 F_A(Q^2)}{M_{\pi}^2 + Q^2}.
\end{equation}
This is the relationship that is
used in all modern neutrino generators\cite{GENIE,NEUT,NEUT2,NUANCE,NuWro}.

$F_V^3$ and $F_A^3$ are form factors associated with the second class
current (SCC). The existence of such currents requires charge or time
symmetry violation, and current measurements show the size of these
violations to be small. Additionally a nonzero $F_V^3$ term would violate
conservation of the vector current (CVC).  Both $F_V^3(0)$ and
$F_A^3(0)$ can be limited experimentally in studies of beta decay.
Almost all current analyses of neutrino data assume 
that the SCCs are zero.  The vector SCCs only enter the cross-section
in terms suppressed by $m^2/M^2$, but there are unsuppressed terms involving
the axial SCC form factor.

\section{Muon and Electron Neutrino Quasi-Elastic Cross Section Differences}

In this section, we will study the dependence of the muon-electron
cross-section differences as a function of $E_\nu$ and $Q^2$.
Differences arise due to the variation of kinematic limits
due to the final state lepton mass, different radiative corrections to
the tree level process and uncertainties in nucleon form factors.
Equations~\ref{eq:Afunc} and \ref{eq:Bfunc} contain explicitly the
dependence of the CCQE cross-section in terms of the form factors.
Lepton mass, $m$, enters in both $A(Q^2)$ and $B(Q^2)$ and these terms
involve all the form factors discussed above.  Note that $F_P$ and
$F_V^3$ {\em only} appear in terms multiplied by $m^2/M^2$ and
therefore are negligible in the electron neutrino cross-section, but
not in the muon neutrino cross-section.  

As metrics, we define the fractional differences
between the muon and electron neutrino CCQE cross-sections:
\begin{eqnarray}
\label{eq:diff}
\delta(E_\nu,Q^2) &\equiv& \frac{\frac{d\sigma_{\mu}}{dQ^2} - \frac{d\sigma_{e}}{dQ^2}}{\int dQ^2\frac{d\sigma_{e}}{dQ^2} } \\
\label{eq:intdiff}
\Delta(E_\nu)&\equiv& \frac{\int dQ^2\frac{d\sigma_{\mu}}{dQ^2} - \int dQ^2\frac{d\sigma_{e}}{dQ^2}}{\int dQ^2\frac{d\sigma_{e}}{dQ^2} }.
\end{eqnarray}
The integrals over $Q^2$ in Eqs.~\ref{eq:diff} and \ref{eq:intdiff}
are taken within the kinematic limits of each process, and those limits
depend on lepton mass as discussed in the next section.

Another useful metric is the difference between a cross-section in a
model with a varied assumption from that of a reference model.  Our
reference model derives $F^1_V$ and $F^2_V$ 
from the electric and magnetic vector
Sachs form factors which follow the dipole form of Eq.~\ref{eq:dipole} with
$C=c^2M_V^2=(0.84)$~(GeV/c)$^2$, and it assumes $F_A$ is a dipole with
$C=c^2M_A^2=(1.1)$~(GeV/c)$^2$.  The reference model uses the derived
$F_P$ from Eq.~\ref{eqn:fp}, and assumes that
$F^3_V=F^3_A=0$ at all $Q^2$.  We then define:
\begin{equation}
\label{eq:flavorDiff}
\Delta_\ell(E\nu)\equiv\frac{\int dQ^2\frac{d\sigma_{\ell}}{dQ^2} -
  \int dQ^2\frac{d\sigma_{\ell}^{ref}}{dQ^2}}{\int
  dQ^2\frac{d\sigma_{\ell}^{ref}}{dQ^2} },
\end{equation}
where $\sigma_\ell^{ref}$ is the reference model for
$\nu_\ell n\to\ell^- p$ or its anti-neutrino analogue and
$\sigma_\ell$ is the model to be compared to the reference.

\subsection{Kinematic Limits}

\begin{figure}[tp]
\centering
    \advance\leftskip-.2in
    \vspace{-10pt}
\begin{tabular}{c}
\includegraphics[width=70mm]{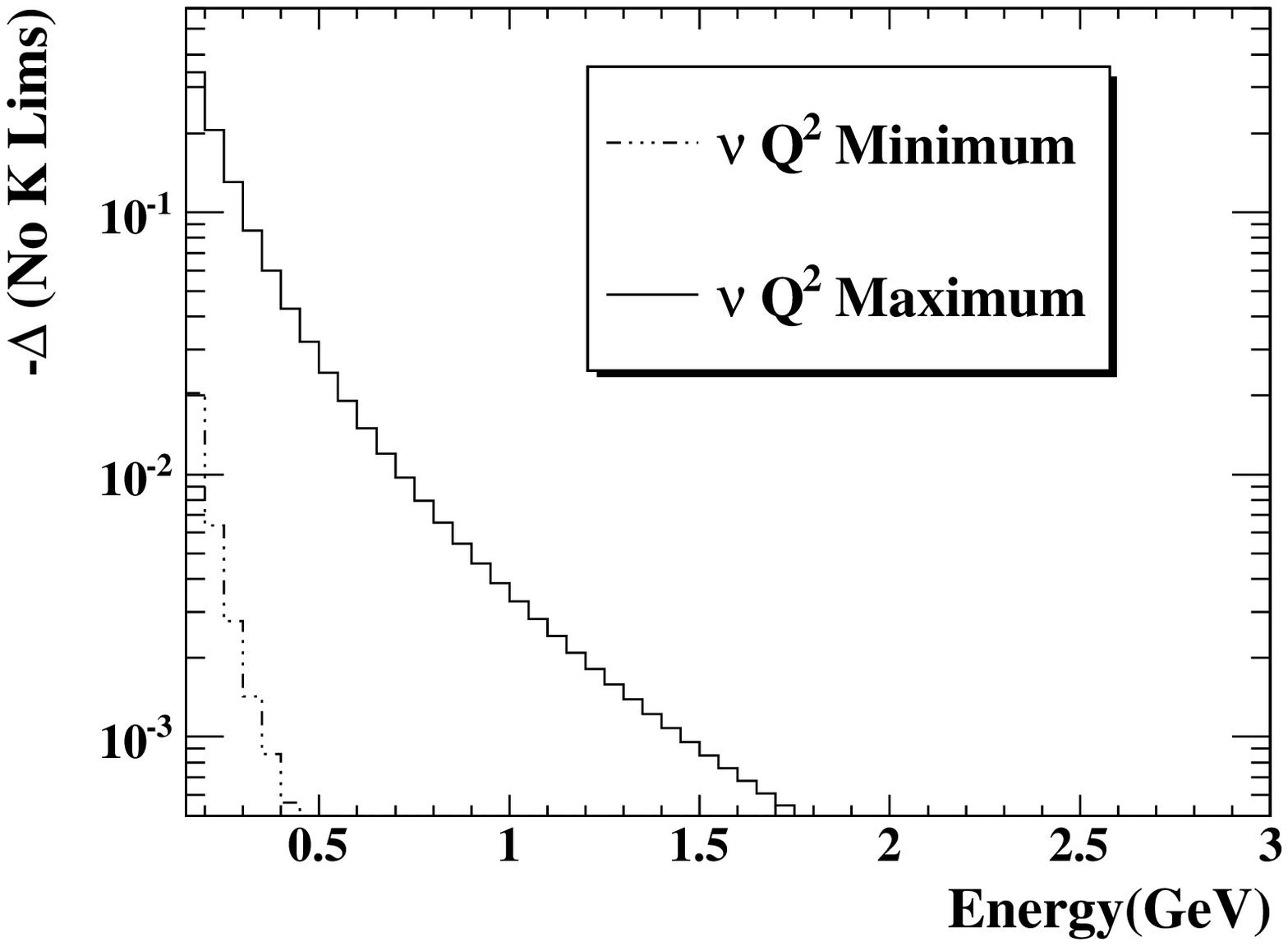} \\
 \includegraphics[width=70mm]{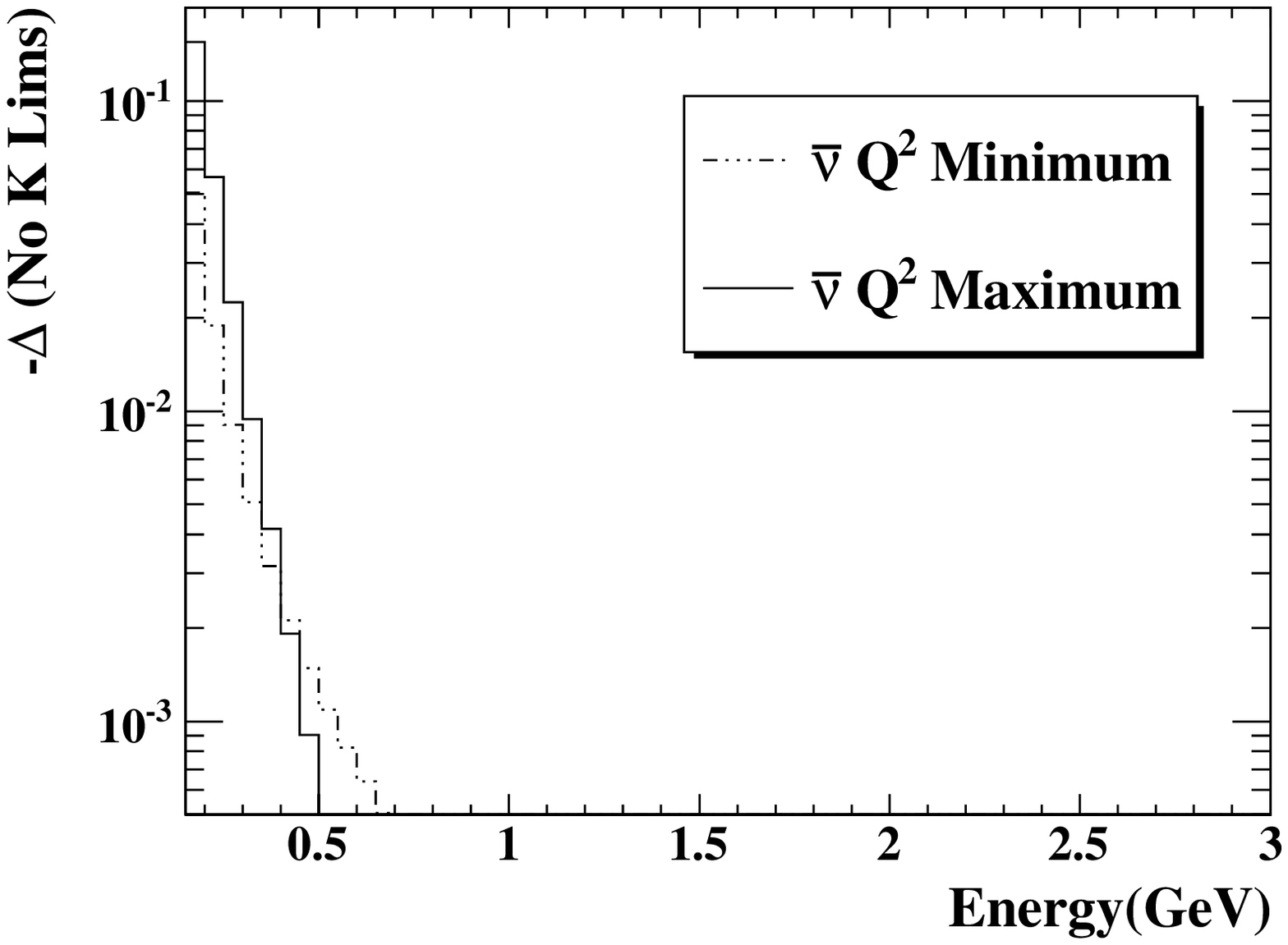} \\
\end{tabular}
    \vspace{-15pt}
\caption{The total charged-current quasi-elastic cross-section
  difference for neutrinos (top) and anti-neutrinos (bottom) due to
  the kinematic limits in $Q^2$.  This difference is $-\Delta$ defined
  in Eq.~\ref{eq:intdiff}, meaning that the electron neutrino
  cross-section is larger than the muon neutrino cross-section.}
    \vspace{-10pt}
\label{fig:kineLim}
\end{figure}

The neutrino and anti-neutrino CCQE processes have kinematic limits in
$Q^2$ which depend on the final state lepton mass, $m$.  These limits
are
\begin{equation}
\label{eq:qmaxmin}
Q^2_{\substack{max\\min}} = - m^2 + \frac{s-M^2}{\sqrt{s}}\left( E^*_\ell \pm |p^*_\ell|  \right)
\end{equation}
where $s$ is the Mandelstam invariant representing total center of
mass energy and $E^*_\ell$ and $p^*_\ell$ are the center of mass final
state lepton energy and momentum.  $E^*_\ell$ can be expressed in
terms of invariants as
\begin{equation}
E^*_\ell = \frac{s+m^2-M^2}{2\sqrt{s}}.
\end{equation}

Figure~\ref{fig:kineLim} shows the effect of the kinematic limits.
Not surprisingly, the effect is very large near the threshold for the
muon neutrino and anti-neutrino reaction.  These effects are accounted
for in the description of the quasi-elastic process in all
commonly used neutrino generators.  However, it is worth noting that
the difference in $Q^2$ spanned by the two reactions can lead to large
effects in varying form factors that significantly affect either the
small or large $Q^2$ parts of the cross-section.
 
\subsection{Radiative Corrections}

\begin{figure}[tp]
\centering
\includegraphics[width=\columnwidth]{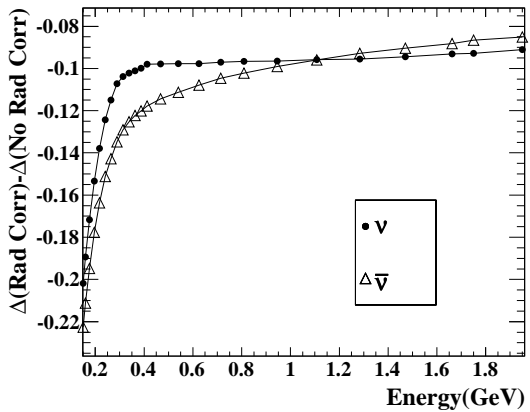}
\caption{Our estimate in the lepton leg leading log approximation 
  of the fractional difference between the electron and muon
  neutrino total charged-current quasi-elastic cross-sections,
  $\Delta$ as defined in Eq.~\ref{eq:intdiff}, as a
  function of neutrino energy.  The negative difference means that 
the electron neutrino
  cross-section is larger than the muon neutrino cross-section.}
\label{fig:radCorr}
\end{figure}

To calculate the effect of radiative corrections on the total
quasi-elastic cross-section, we follow the approximate approach of
calculating the leading log correction to order $\log Q/m$, where $Q$
is the energy scale of the interaction process\cite{De_Rujula:1979jj}.  
This approach has a
calculational advantage in investigating the differences due to the
lepton mass, $m$ because the lepton leg leading log only involves
sub-processes where photons attach to leptons.
The key result from this approach is that the cross-section which
allows for the presence of radiated photons, $\sigma_{LLL}$ is related
to the Born level cross-section, $\sigma_B$, by
\begin{eqnarray}
\label{eqn:deRlll}
\frac{d\sigma_{LLL}}{dE_\ell d\Omega}&\approx&\frac{d\sigma_B}{dE_\ell
  d\Omega}+\frac{\alpha_{EM}}{2\pi}\log\frac{4E^*_\ell}{m^2}\int^1_0dz\frac{1+z^2}{1-z}
\nonumber\\
&&\times\left( \frac{1}{z}\frac{d\sigma_B}{d\hat{E}_\ell
  d\Omega}\left| _{\hat{E}_\ell=E_\ell/z} \right. -\frac{d\sigma_B}{dE_\ell
  d\Omega}\right) ,
\end{eqnarray}
where $E^*_\ell$ is the center-of-mass frame lepton energy.  

In the
case of elastic scattering, the relationship in
$\sigma_B$ between $E_\ell$ and the scattering angle, $\theta_\ell$ simplifies
the calculation because there is at most one $z$ in the integrand for
which the cross-section does not vanish for a particular lepton angle:
\begin{eqnarray}
z&&=\left[2 E_{\ell } \left(M+E_{\nu }\right)
   \left(m^2+2 M E_{\nu }\right) -2\cos^2\theta _{\ell}E_{\ell }E_{\nu
   }\right.\nonumber\\
&&\times\left.\sqrt{m^4+4
   E_{\nu }^2 \left(M^2-m^2\sin^2\theta_{\ell}\right)-4 m^2 M^2-4 m^2 M
   E_{\nu }}\right] \nonumber\\
&&/\left[m^4+4 E_{\nu } \left(E_{\nu }
   \left(m^2 \cos ^2\theta _{\ell}+M^2\right)+m^2 M\right)\right] .
\end{eqnarray}
We then obtain the
remaining cross-section by integrating
Eq.~\ref{eqn:deRlll} over the final state lepton energy.  Note that
this procedure only gives a prescription for evaluating
$d\sigma(E_{\nu\text , true})/dQ^2_{\text true}$; however, the
radiation of real photons means that the relationship between
lepton energy and angle and $E_\nu$ and $Q^2$ in elastic scattering
will no longer be valid.  The effect of this distortion of the elastic
kinematics will depend on the details of the experimental
reconstruction and the neutrino flux seen by the experiment, so the
effect must be evaluated in the context of a neutrino interaction
generator and full simulation of the reconstruction for a given
experiment.

The difference of the effect on the total cross-sections as a function
of neutrino energy is shown in Fig.~\ref{fig:radCorr}.  We estimate
a difference of approximately 10\% over the energies of interest in
oscillation experiments.  The largest differences fractional
differences in cross-sections are at high
true $Q^2$ and low neutrino energies. The magnitude of the lepton leg correction to the muon
neutrino total cross-section is smaller, roughly $0.4$ times this
difference, so the larger effect is on the electron neutrino cross-section.

Our estimation of the effect is surprisingly large at the relevant energies for
oscillation experiments.  Some portion of this difference in the total
cross-section in Fig.~\ref{fig:radCorr} may
be canceled by diagrams missing from the leading log correction in
the lepton leg, such as box diagrams involving $W\gamma$ exchange
between the leptonic legs and the initial or final state, which will
also depend on the final state lepton mass~\cite{Sirlin:1981yz}.
We stress that this is only an
approximate treatment which should be confirmed in a full calculation
implemented inside a generator, and to date radiative corrections are 
not included in the commonly used
neutrino interaction generators\cite{GENIE,NEUT,NEUT2,NUANCE}.

\subsection{Uncertainties in $F^1_V$, $F^2_V$ and $F_A$} 

\begin{figure}[tp]
\centering
    \advance\leftskip-.2in
\includegraphics[width=70mm]{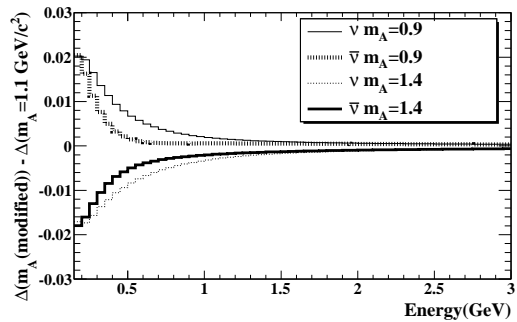}
\caption{The change in the fractional difference of muon CCQE
  cross-section and electron CCQE when $m_A$ is changed from a
  reference value of $1.1$~GeV/c$^2$ in a range generously consistent with
  current experimental data.}
    \vspace{-10pt}
\label{fig:mA}
\end{figure}

As noted above, the vector form factors $F_V^1$ and $F_V^2$ are
precisely measured in charged lepton scattering\cite{BBBA}; however,
the axial form factor is still uncertain because neutrino experiments
that measure it do not agree amongst themselves or with determinations
in pion electroproduction as discussed above.  Therefore the axial
form factor will dominate any differences in the electron and muon
cross-sections due to uncertainties in leading form factors.

Figure~\ref{fig:mA} illustrates the change in the fractional
difference of muon and electron neutrino CCQE cross-sections when the
axial form factor is varied by changing the assumed dipole mass in a
range consistent with experimental measurements.  The size of the effect is of order
1\% at very low energy and drops with increasing energy.  This
difference in cross-section may be accounted for in variations
of the axial form factor within the analysis of an experiment using a
modern neutrino interaction generator.

\subsection{Pseudoscalar Form Factor}


At low $Q^2$, the pseudoscalar form factor does have a significant
contribution to the muon neutrino CCQE cross-section, of nearly the
same order of the leading terms.  However, Eq.~\ref{eqn:fp} shows
that the contribution will be suppressed for
$Q^2\stackrel{>}{\sim}M_\pi^2$, and all terms involving $F_P$ are
suppressed by $m/M$ and so the contribution to the cross-section is
negligible for electron neutrinos.  At low neutrino energies, the
pseudoscalar form factor effect on the cross-section difference,
$\Delta(E_\nu)$ is nearly as large as that of the kinematic limits.
The effect of the form factor as a function of neutrino energy and
$Q^2$ is different for neutrinos and anti-neutrinos. 

\begin{figure}[tp]
    \vspace{-12.2pt}
\includegraphics[width=\columnwidth]{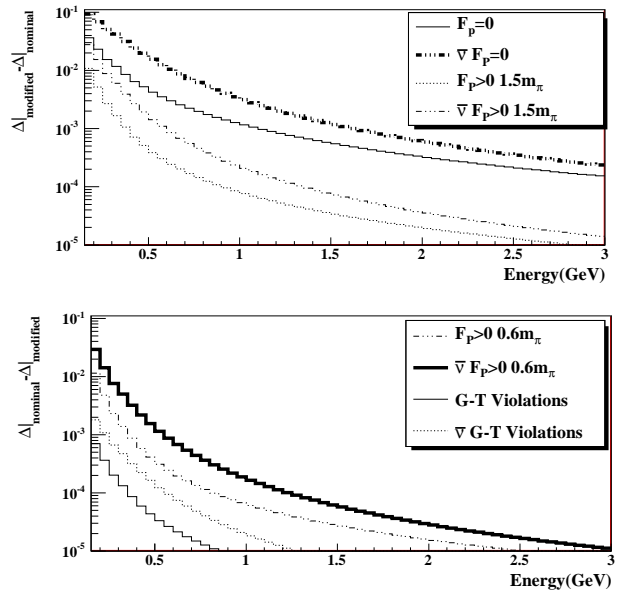} 
    \vspace{-15pt}
\caption{The effect of variations of $F_P$ from the reference model
  which assumes PCAC and the Goldberger-Treiman relation.  The plots
  illustrate the change cross-section difference, $\Delta(E_\nu)$,
  between a varied model and the reference model.  Possible violations
  of the G-T relation produce a negligibly small effect, even at low
  energy.  The range of violations from PCAC allowed by current data
  would allow significantly larger changes.  The effect of setting $F_P$ to
  zero is shown for comparison.}
    \vspace{-13pt}
\label{fig:fpUncs}
\end{figure}

Current neutrino interaction generators\cite{GENIE,NEUT,NEUT2,NUANCE}
include the effect of $F_P$ shown in Eq.~\ref{eqn:fp} under the
assumptions of PCAC and that the Goldberger-Treiman relation holds for
all $Q^2$.  Experimental tests of the Goldberger-Treiman relation have
identified small discrepancies which imply that the left hand side of
Eq.~\ref{eq:gt} is between 1\% and 6\% less than the right-hand
side\cite{gtlim,gtrange}.  Guidance from models suggests that this
effect is likely to disappear at high $Q^2$\cite{gtq2}.  We examine
the effect of varying $F_P(0)$ by 3$\%$ of itself as a reasonable
approximation to the possible difference due to this effect.  A more
significant difference may arise due to violations of PCAC.  This has
been directly checked in pion electroproduction studies\cite{Choi:1993vt}
which can directly measure $F_P(Q^2)$ in the range of $0.05$ to
$0.2$~GeV/c$^2$.  Uncertainties in this data limit the reasonable
range of pole masses in Eq.~\ref{eqn:PCACderiv} to be 
between $0.6M_\pi$ and $1.5M_\pi$.  Effects due to these
possible deviations from PCAC and the Goldberger-Treiman relation are
shown in Fig.~\ref{fig:fpUncs} along with the effect of assuming
$F_P=0$ for comparison.

\subsection{Second Class Currents}

As noted in the introductory material, non-zero second class currents
violate a number of symmetries and hypotheses, and are therefore
normally assumed to be zero in analysis of neutrino reaction data
and in neutrino interaction generators.  For this study, we take a data
driven approach and look at the effect of the largest possible
second-class current form factors, $F_V^3$ and $F_A^3$ that do not
violate constraints from this data.  

Vector second-class currents
enter the cross-sections for neutrino quasi-elastic scattering always
suppressed by $m/M$ and therefore only appear practically in muon
neutrino scattering cross-sections.  Both vector and axial vector form
factors give large contributions to the $B(Q^2)$ term given in
Eqs.~\ref{eq:lsshort} and \ref{eq:Bfunc}, and therefore typically
have very different effects, often even different in sign, for
neutrino and anti-neutrino scattering.

\begin{figure}[tp]
\centering
\begin{tabular}{c}
\includegraphics[width=70mm]{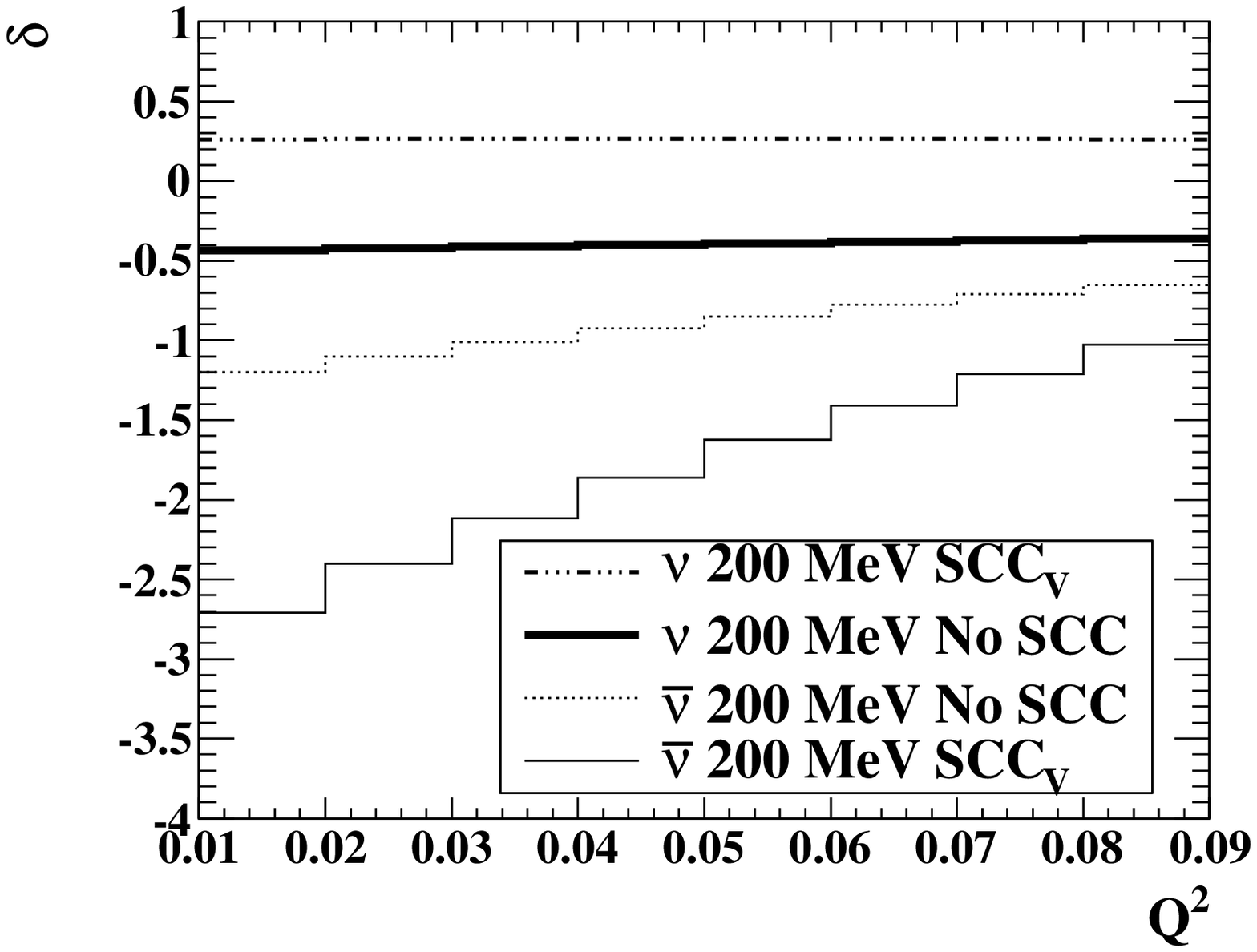} \\
\includegraphics[width=70mm]{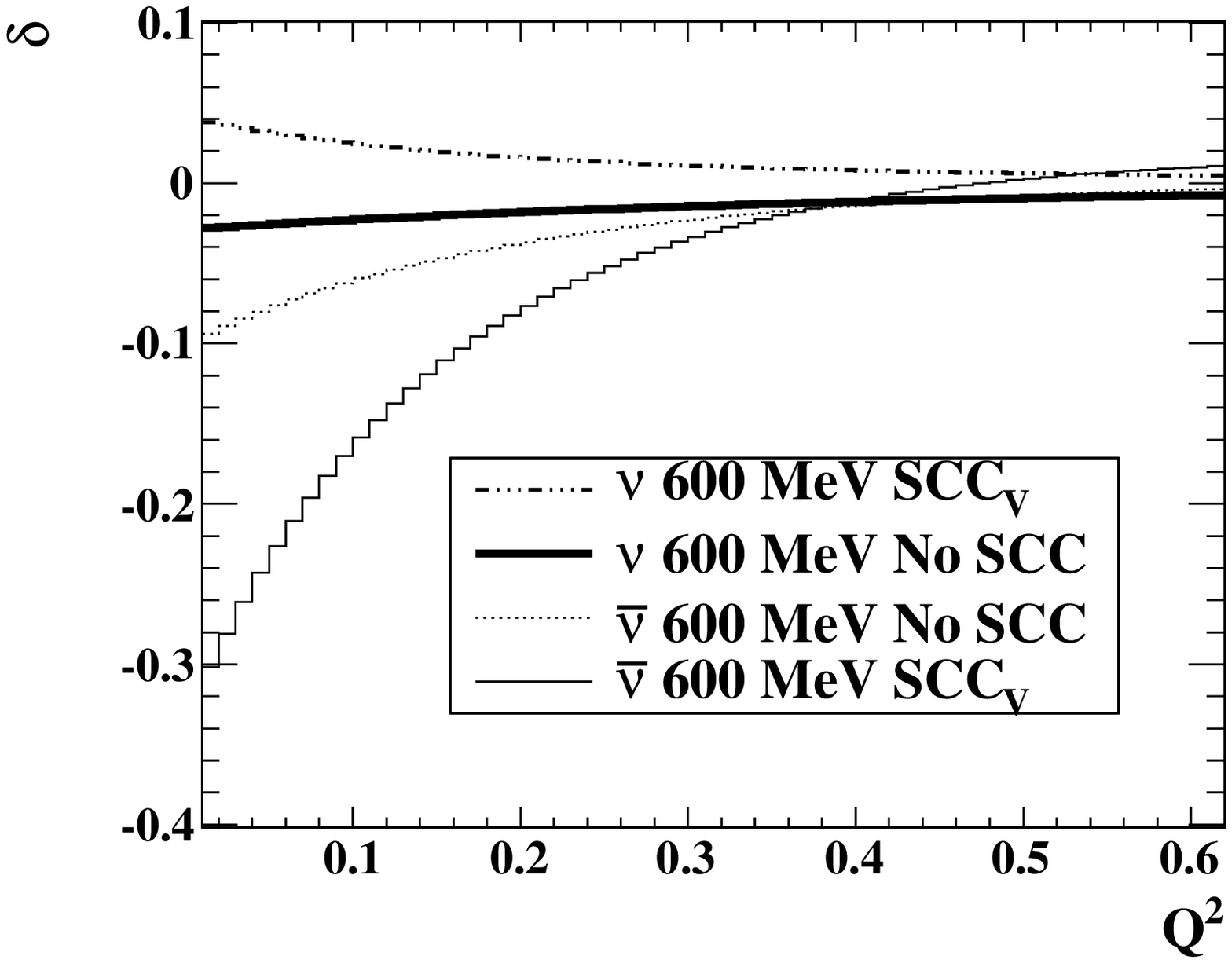} \\
\includegraphics[width=70mm]{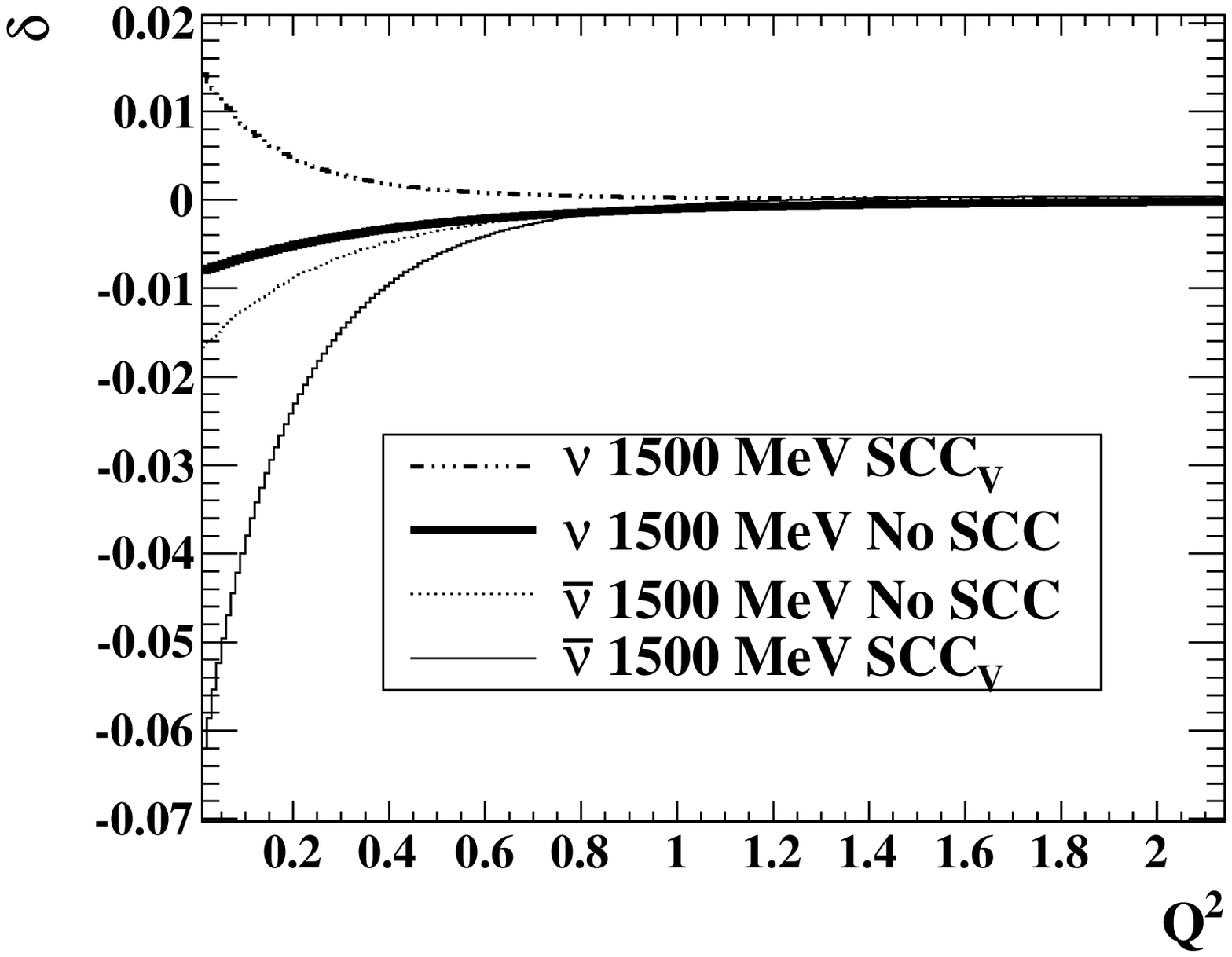} \\
\end{tabular}
\caption{$\delta(E_\nu,Q^2)$, defined in Eq.~\ref{eq:diff}, as a function of $Q^2$ for several selected $E_\nu$.  The difference between including and not including the maximum allowed second class vector current (``SCC$_{\text V}$''), $F^3_V(Q^2)=4.4F^1_V(Q^2)$, is shown.}
\label{fig:q2fvdiff}
\end{figure}


\begin{figure}[tp]
\centering
\includegraphics[width=70mm]{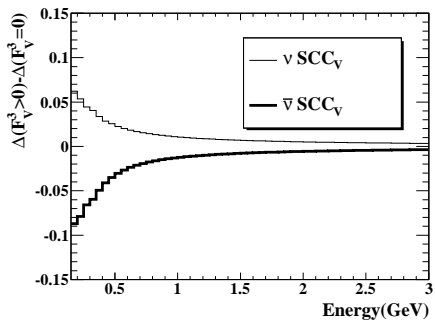} 
\caption{Changes in the difference between the muon and electron
  neutrino cross sections due to including $F_V^3$.}
\label{fig:threecon2}
\end{figure}

The vector second-class currents are difficult to detect in most
weak processes involving electrons because the process is generally
suppressed by powers of $m_e/M$.  Therefore even very precise beta
decay measurements have difficulty limiting the size of $F_V^3(0)$ to
less than several times the magnitude of the regular vector form
factors\cite{Holstein:1984ga}.  The best limits from beta decays
currently limit $F^3_V(0)/F^1_V(0)$ to be $(0.0011\pm
0.0013)\frac{m_N}{m_e}\approx2.0\pm2.4$\cite{Hardy:2004id}.  Studies
of muon capture on nuclei can provide modestly better limits, but at
the expense of assuming there are no axial second class
currents\cite{Holstein:1984ga}.  An analysis of anti-muon neutrino
quasi-elastic scattering has been used to place limits of similar
strength, but again under the assumption of no axial second class
currents and with an assumed $Q^2$ dependence,
$F_V^3(Q^2)=F_V^3(0)/(Q^2+M^2_{3V})^2$ with a fixed $M_{3V}$ of
$1.0~GeV/c^2$\cite{Ahrens:1988rr}.  From the preponderance of the
data, we choose to parameterize the maximum size of the allowed vector
second class current as $F^3_V(Q^2)=4.4F^1_V(Q^2)$, which is
not excluded by the results of any of the above
studies.  The effect of this is significant, particularly at low
neutrino energies and is shown in Figs.~\ref{fig:q2fvdiff} and
\ref{fig:threecon2}. Recall that the effect on the electron neutrino
cross-section from $F^3_V$ is negligible, so this effect occurs
almost entirely in the muon neutrino cross-section.

\begin{figure}[tp]
\centering
\includegraphics[width=70mm]{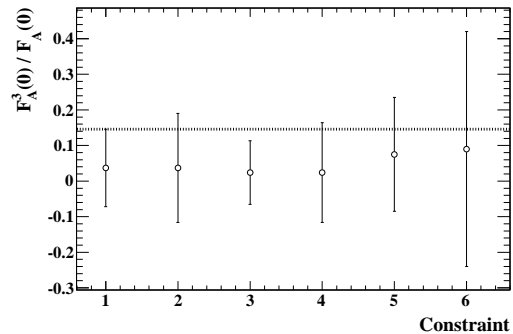} 
    \vspace{-15pt}
\caption{A survey of constraints on the ratio $F_A^3(0)$/$F_A(0)$ with
  their uncertainties from: (1) Wilkinson's data
  compilation\cite{wilkexp}, (2) the same Wilkinson compilation with a
  correction for short-range effects\cite{wilkexp}, (3) the method of
  Wilkinson applied only to the $A=20$ KDR
  parameters\cite{wilkexp,a20}, (4) ibid, with a correction for
  short-range effects\cite{wilkexp,a20}, (5) a derived limit from
  $A=12$ beta decays\cite{a12} and (6) a derived limit from $A=20$
  beta decays\cite{a20}.  The value used for $F_A^3(0)/F_A(0)$ in
  this study is shown by the dashed line.}
    \vspace{-13pt}
\label{fig:scccon}
\end{figure}
By contrast, the axial second class current at zero $Q^2$ is reasonably well
constrained by studies of beta decay.  We derive our limits in the
framework of the KDR parameters\cite{kubodera} where there is a wealth
of experimental data to constrain these
parameters\cite{kubexp,wilkexp,a12,a20} and therefore derive a limit
on $F_A^3(0)$.  Figure~\ref{fig:scccon} shows these experimental
constraints and the effect we allow in this study.  


\begin{figure}[tp]
\centering
\begin{tabular}{c}
\includegraphics[width=70mm]{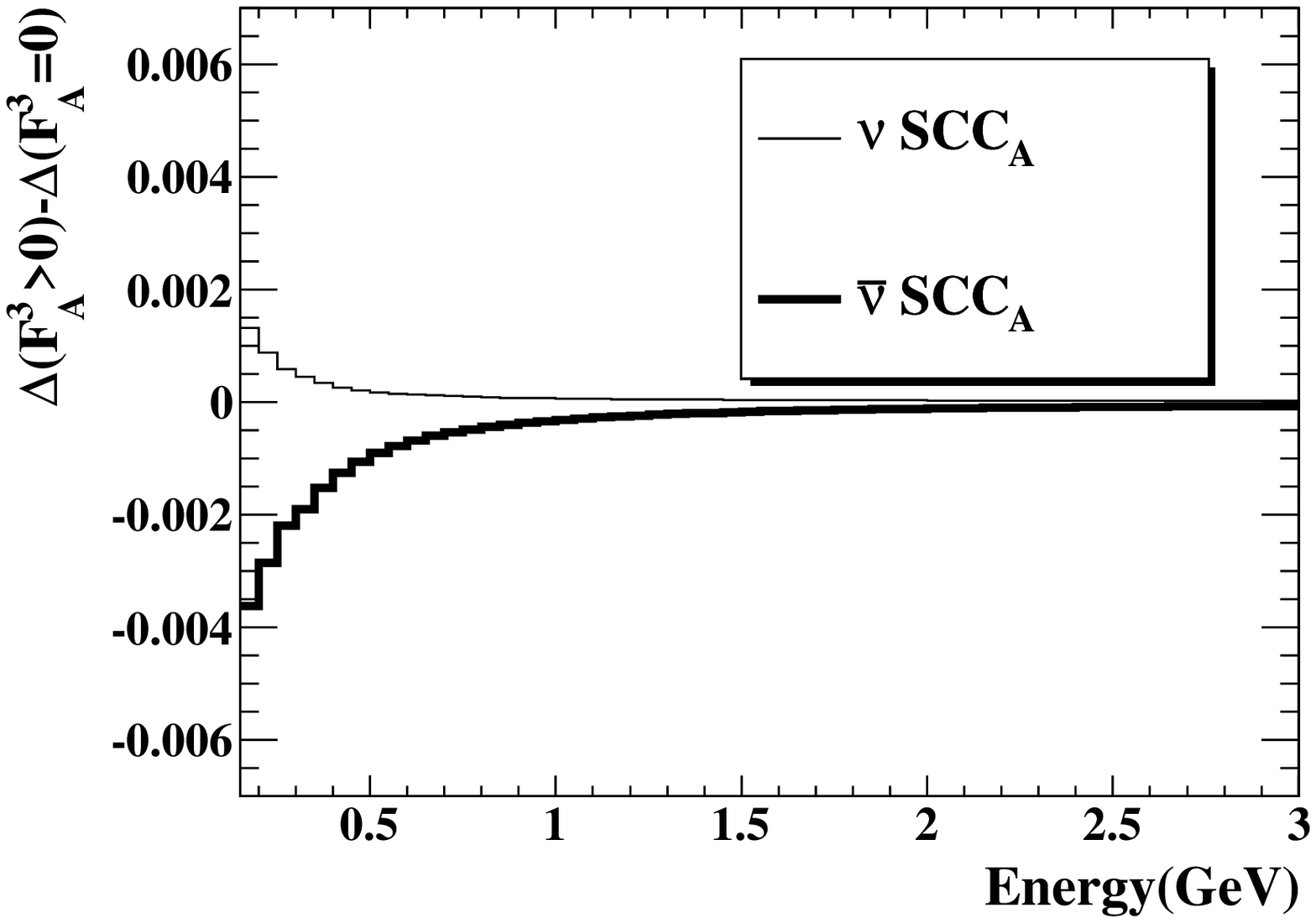} \\
\includegraphics[width=70mm]{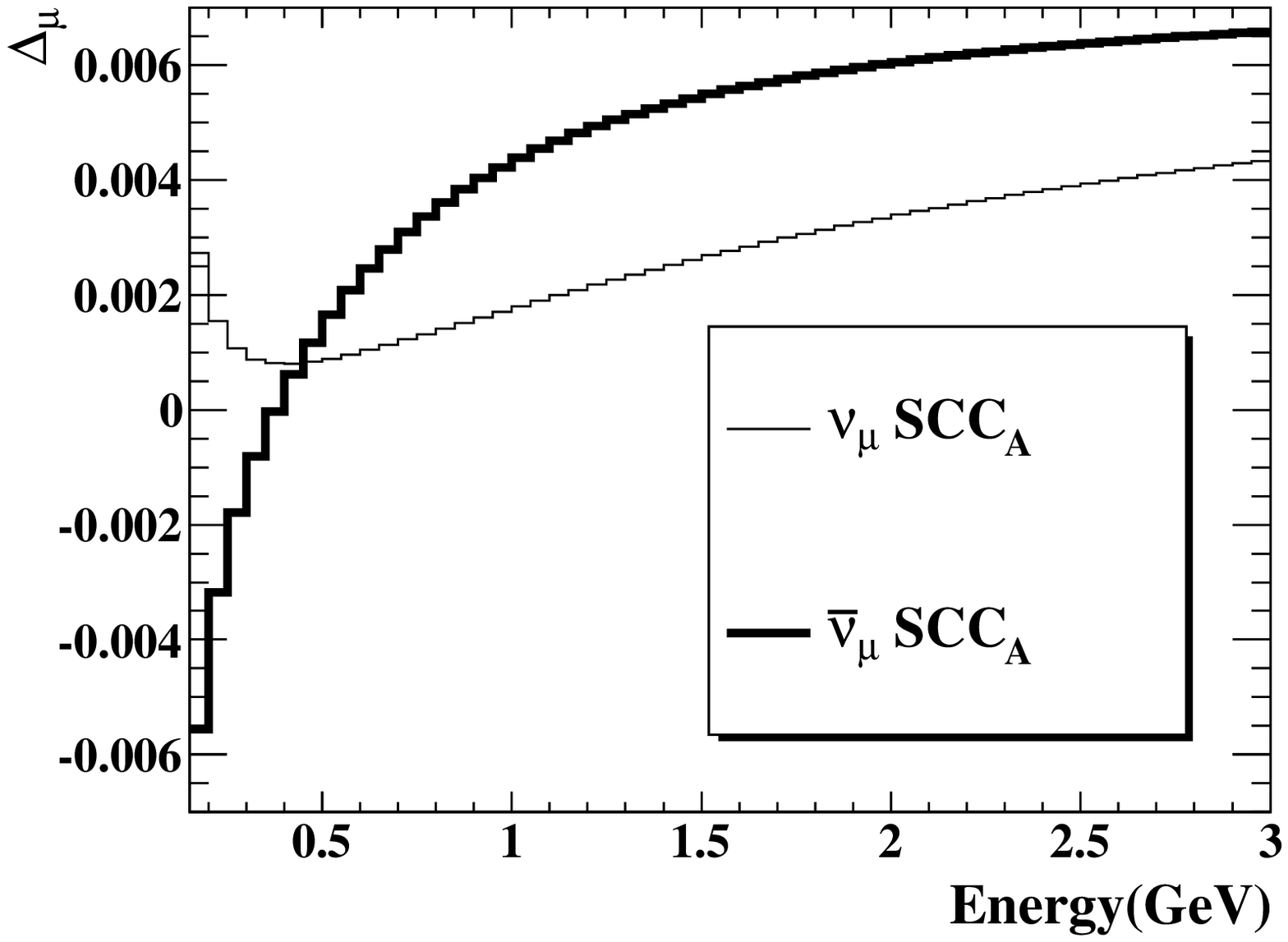} 
\end{tabular}
\caption{Top: Changes in the difference between the muon and electron neutrino cross sections due to including $F_A^3$; Bottom: the change in muon neutrino cross-sections due to including $F_A^3$.
}
\label{fig:threecon}
\end{figure}

We assume a dipole form for the variation in $Q^2$ of the axial second
class current as well, so that for the maximum allowed variation
$F_A^3(Q^2)/F_A(Q^2)=F_A^3(0)/F_A(0)=0.15$.  Figure~\ref{fig:threecon}
shows the effect of including this allowed axial second class current
on both the difference of electron and muon neutrino cross-sections
and on the muon neutrino cross-section itself.
It is significantly smaller than the effect of the vector second class
current because the limits on these currents are more stringent. 

\section{Conclusions}

\begin{figure}[tp]
\centering
\begin{tabular}{c}
\includegraphics[width=70mm]{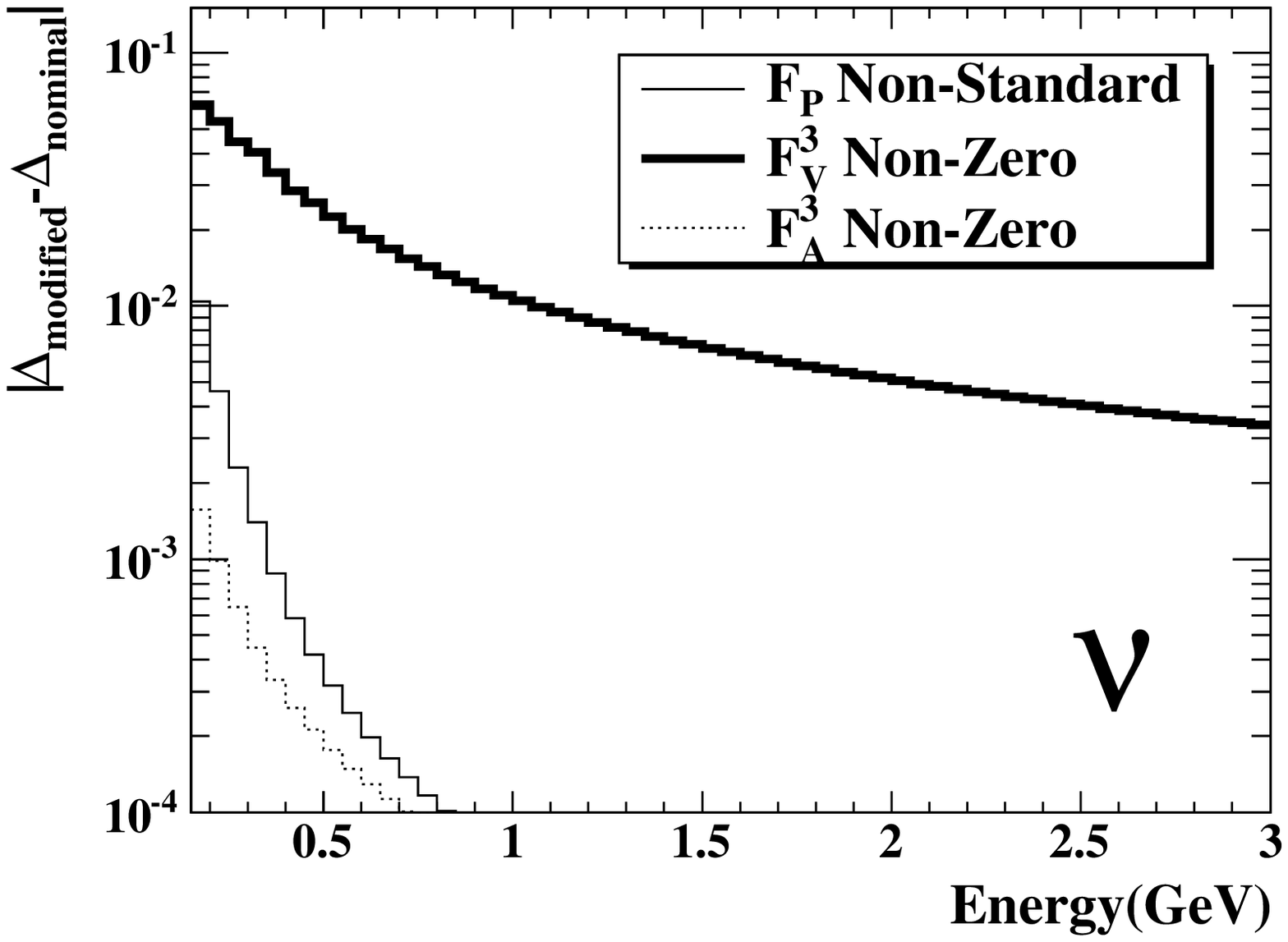} \\
\includegraphics[width=70mm]{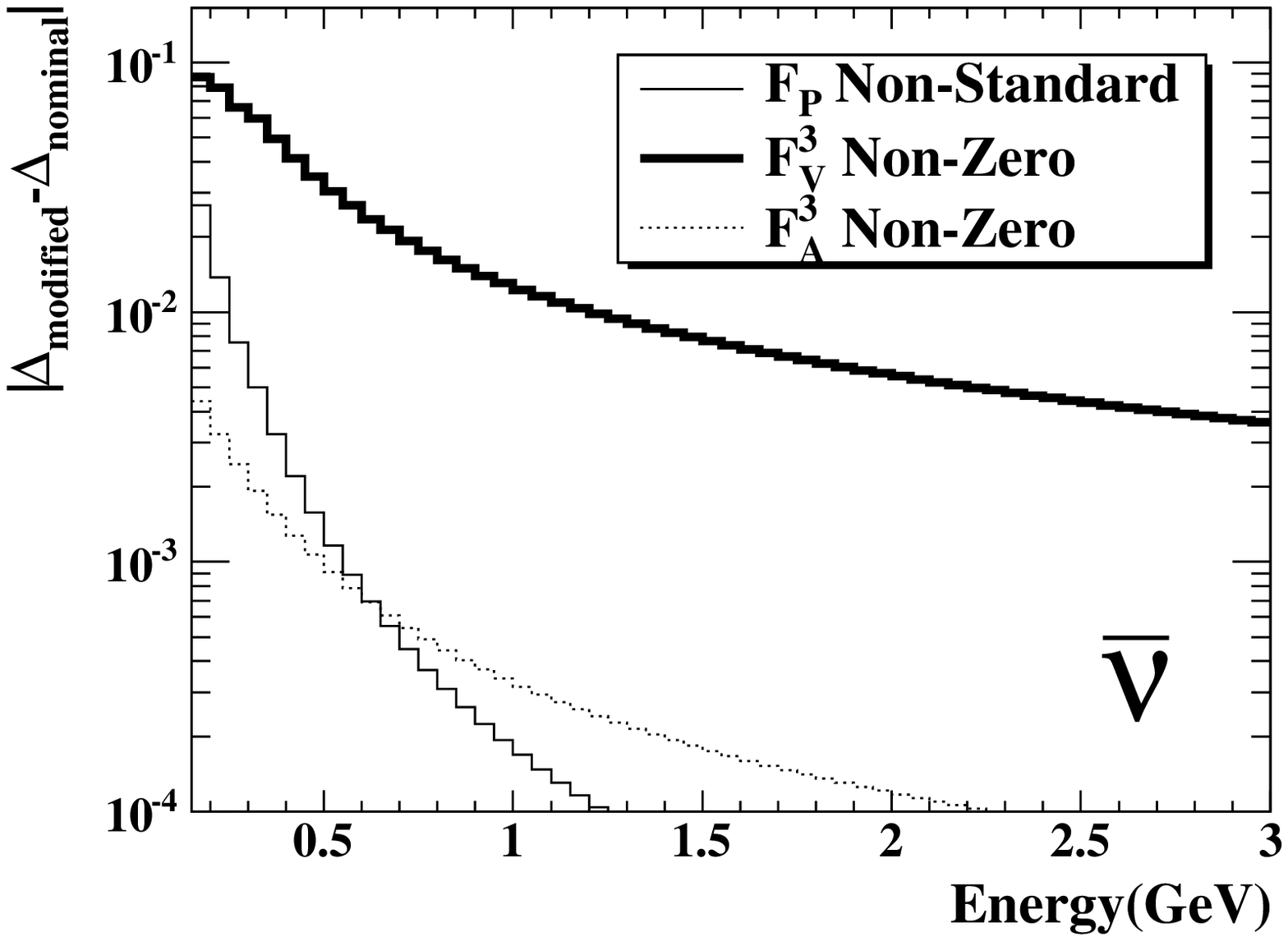}\\ 
\includegraphics[width=70mm]{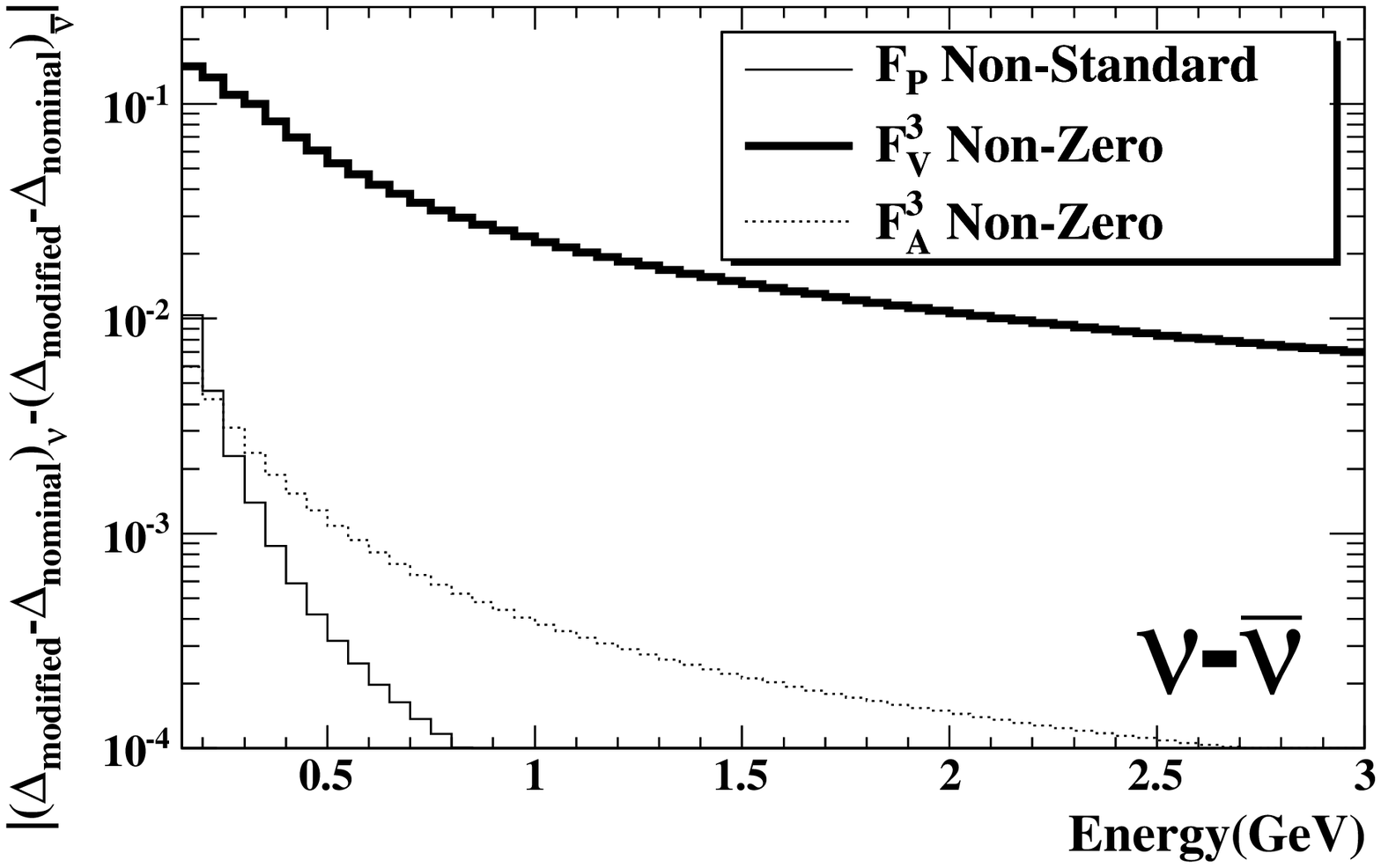}\\ 
\end{tabular}
\caption{Top and Middle: For the form factors not well constrained and not
  accounted for in neutrino generators, a summary of the magnitude of the fractional size of
  differences in the total
  charged-current quasi-elastic cross-sections between electron and
  muon neutrinos and anti-neutrinos as a function of neutrino energy.
  For $F_{P}$ the average of the magnitude of the PCAC violating effects are 
  summed linearly with the magnitude of the Goldberger-Treiman violation effect.
  Bottom: The magnitude of the difference between $\nu$ and $\bar{\nu}$ of the fractional
  differences which illustrates the size of apparent CP violating
  asymmetries in oscillation experiments.}
\label{fig:summary}
\end{figure}

Large differences between the electron and muon neutrino quasi-elastic
cross-sections exist at low neutrino energies from the
presence of different kinematic limits due to the final state lepton
mass and due to the presence of the pseudoscalar form factor, $F_P$,
derived from PCAC and the Goldberger-Treiman relation.  These
differences are typically accounted for in modern neutrino interaction
generators.

There are also significant differences due to radiative
corrections, particularly in diagrams that involve photon radiation
attached to the outgoing lepton leg which are proportional to
$\log Q/m$.  These differences are calculable, but are
typically not included in neutrino interaction generators employed by
neutrino oscillation experiments.  If our estimate of these
differences, of order $10\%$, is confirmed by more complete analyses,
then this is a correction that needs to be included
as it is comparable to the size of current systematic uncertainties at
accelerator experiments\cite{T2K1,T2K2}.

Modifications of the assumed $F_P$ from PCAC and the
Goldberger-Treiman relation and the effect of the form factors $F_V^3$
and $F_A^3$ corresponding to second class vector and axial currents,
respectively, are not included in neutrino interaction generators.  A
summary of the possible size of these effects, as we have estimated
them, is shown in Fig.~\ref{fig:summary}.

These differences, particularly from the 
second class vector currents, may
be significant for current\cite{T2K1,T2K2,NOvA} and future\cite{LBNE}
neutrino oscillation experiments which seek precision measurements of
$\nu_\mu\to\nu_e$ and its anti-neutrino counterpart at low neutrino
energies.  Previous work\cite{Ahrens:1988rr} has
demonstrated sensitivity to these second class currents in neutrino
and anti-neutrino quasi-elastic muon neutrino scattering, and future
work with more recent
data\cite{Lyubushkin:2008pe,AguilarArevalo:2010zc} and newly analyzed
data\cite{MINERvA-QE} may help to further limit uncertainties on
possible second class currents.

\begin{acknowledgments}
The suggestion for this work came out of conversations with Alain
Blondel about systematics in future oscillation experiments and we
thank him for inspiring this work.  We are grateful to Ashok Das,
Tamar Friedmann and Tom McElmurry for their clear and patient
explanations of the bilinear covariant structure of weak interactions.
We thank Arie Bodek for a helpful discussion of available tests of the
CVC hypothesis.  We thank Gabriel Perdue and Geralyn Zeller for helpful
comments on a draft of this manuscript.  We are grateful to Bill
Marciano for his helpful insights into the radiative corrections after the initial
draft of this paper appeared online.

This material is based upon work supported by the Department of Energy
under Award Number DE-FG02-91ER40685.

\end{acknowledgments}


\begin{thebibliography}{5}

\bibitem{Lederman-Schwartz-Steinberger} G.~Danby {\em et al.},
Phys.\ Rev.\ Lett.\ {\bf 9}, 36--44 (1962).

\bibitem{T2K1} 
Y.~Itow {\it et al.}  [The T2K Collaboration],
hep-ex/0106019.
\bibitem{T2K2} 
  K.~Abe {\it et al.}  [T2K Collaboration],
  Phys.\ Rev.\ Lett.\  {\bf 107}, 041801 (2011).

\bibitem{NOvA} D. Ayres {\em et al.} [NOvA Collaboration], 
hep-ex/0503053 (2004).

\bibitem{De_Rujula:1979jj} 
  A.~De Rujula, R.~Petronzio and A.~Savoy-Navarro,
  Nucl.\ Phys.\ B {\bf 154}, 394 (1979).

\bibitem{GENIE} 
  C.~Andreopoulos [GENIE Collaboration],
  Acta Phys.\ Polon.\ B {\bf 40}, 2461 (2009).

\bibitem{NEUT}  Y.~Hayato,
  Nucl.\ Phys.\ Proc.\ Suppl.\  {\bf 112}, 171 (2002).
\bibitem{NEUT2} 
  Y.~Hayato,
  Acta Phys.\ Polon.\ B {\bf 40}, 2477 (2009).
\bibitem{NUANCE}  D.~Casper,
  Nucl.\ Phys.\ Proc.\ Suppl.\  {\bf 112}, 161 (2002).

\bibitem{Smith-Moniz}R.A. Smith and E.J. Moniz
Nucl.\ Phys.\ {\bf B43} 605 (1972).

\bibitem{Bodek-Ritchie}A. Bodek and J.L. Ritchie,
Phys.\ Rev.\ {\bf D23} 1070 (1981).

\bibitem{Benhar} 
O.~Benhar, A.~Fabrocini, S.~Fantoni and I.~Sick,
Nucl.\ Phys.\ A {\bf 579}, 493 (1994).

\bibitem{NuWro}
J.~Sobczyk,
PoS NUFACT {\bf 08}, 141 (2008).

\bibitem{marshak}R.E~Marshak, Riazuddin and C.P.~Ryan, 
\textsl{Theory of Weak Interactions in Particle Physics},
Wiley-Interscience (1969). 

\bibitem{lsmith}C.H.~Llewellyn-Smith,
Phys\. Rept. {\bf 3C}, 261--379 (1972).

\bibitem{wilkinson}D.H.~Wilkinson,
Nucl.\ Inst.\ \& Meth.\ {\bf A455}, 656--659 (2000).

\bibitem{Sirlin:1981yz} 
  A.~Sirlin and W.~J.~Marciano,
  Nucl.\ Phys.\ B {\bf 189}, 442 (1981).

\bibitem{BBBA}A.~Bodek, S.~Avvakumov, R.~Bradford and H.~Budd,
Eur.\ Phys.\ J. {\bf C53}, 349--354 (2008).

\bibitem{Bodek:2007vi} 
  A.~Bodek, S.~Avvakumov, R.~Bradford and H.~S.~Budd,
  J.\ Phys.\ Conf.\ Ser.\  {\bf 110}, 082004 (2008).
\bibitem{Lyubushkin:2008pe} 
  V.~Lyubushkin {\it et al.}  [NOMAD Collaboration],
  Eur.\ Phys.\ J.\ C {\bf 63}, 355 (2009).
\bibitem{AlcarazAunion:2009ku} 
  J.~L.~Alcaraz-Aunion {\it et al.}  [SciBooNE Collaboration],
  AIP Conf.\ Proc.\  {\bf 1189}, 145 (2009).
\bibitem{Dorman:2009zz} 
  M.~Dorman [MINOS Collaboration],
  AIP Conf.\ Proc.\  {\bf 1189}, 133 (2009).
\bibitem{AguilarArevalo:2010zc}
  A.~A.~Aguilar-Arevalo {\it et al.}  [MiniBooNE Collaboration],
  Phys.\ Rev.\  D {\bf 81}, 092005 (2010).

\bibitem{Choi:1993vt} 
  S.~Choi, V.~Estenne, G.~Bardin, N.~De Botton, G.~Fournier, P.~A.~M.~Guichon, C.~Marchand and J.~Marroncle {\it et al.},
  Phys.\ Rev.\ Lett.\  {\bf 71}, 3927 (1993).
\bibitem{Liesenfeld:1999mv}
  A.~Liesenfeld {\it et al.}  [A1 Collaboration],
  Phys.\ Lett.\  B {\bf 468}, 20 (1999).

\bibitem{adler}Stephen L. Adler,
Phys.\ Rev.\ {\bf 137}, 1022--1033 (1964).
\bibitem{goldtrei}M. L. Goldberger and S. B.Treiman,
Phys.\ Rev.\ {\bf 5}, 1178-1184 (1958).

\bibitem{gtrange} Thomas Becher and Heinrich Leutwyler,
JHEP {\bf 6}, 17-34 (2001).

\bibitem{gtlim} Jos\'{e} L. Goitya, Randy Lewisa, Martin Schvellingera and Longzhe Zhanga,
Phys.\ Lett.\ {\bf B454}, 115–122 (1999).

\bibitem{gtq2} C.~Alexandrou, G.~Koutsou, Th.~Leontiou, J.W.~Negele
  and A.~Tsapalis,
Phys.\ Rev.\ {\bf D76}, 094511 (2007).




\bibitem{Holstein:1984ga} 
  B.~R.~Holstein,
  Phys.\ Rev.\ C {\bf 29}, 623 (1984).

\bibitem{Hardy:2004id} 
  J.~C.~Hardy and I.~S.~Towner,
  Phys.\ Rev.\ C {\bf 71}, 055501 (2005)
  [nucl-th/0412056].

\bibitem{Ahrens:1988rr} 
  L.~A.~Ahrens {\it et al.},
  Phys.\ Lett.\ B {\bf 202}, 284 (1988).





\bibitem{kubodera} K. Kubodera, J. Delorme and M. Rho,
Nucl. Phys. {\bf B66}, 253-292 (1973).
\bibitem{kubexp} M.~Oka and K.~Kubodera, 
Phys.\ Lett.\ {\bf B90} 45 (1980).
\bibitem{a12}K.~Minamisono {\em et al.},
Phys.\ Rev.\ {\bf C65}, 015501 (2001).
\bibitem{a20}K.~Minamisono {\em et al.}, 
Phys.\ Rev.\ {\bf C84}, 055501 (2011).
\bibitem{wilkexp} D.H.~Wilkinson,
Eur.\ Phys.\ J.\ {\bf A7} 307 (2000).


\bibitem{LBNE} 
  T.~Akiri {\it et al.}  [LBNE Collaboration],
  arXiv:1110.6249 [hep-ex].

\bibitem{MINERvA-QE} 
  K.~S.~McFarland [The MINERvA Collaboration],
  AIP Conf.\ Proc.\  {\bf 1405}, 95 (2011)
  [arXiv:1108.0702 [hep-ex]].
\end{thebibliography}
\end{document}